\documentclass[10pt,letterpaper]{article}
\usepackage[top=0.85in,left=2.75in,footskip=0.75in]{geometry}

\usepackage{amsmath,amssymb}

\usepackage{changepage}

\usepackage{textcomp,marvosym}

\usepackage{cite}

\usepackage{nameref,hyperref}

\usepackage{pdfpages}
\def\supplementfilename{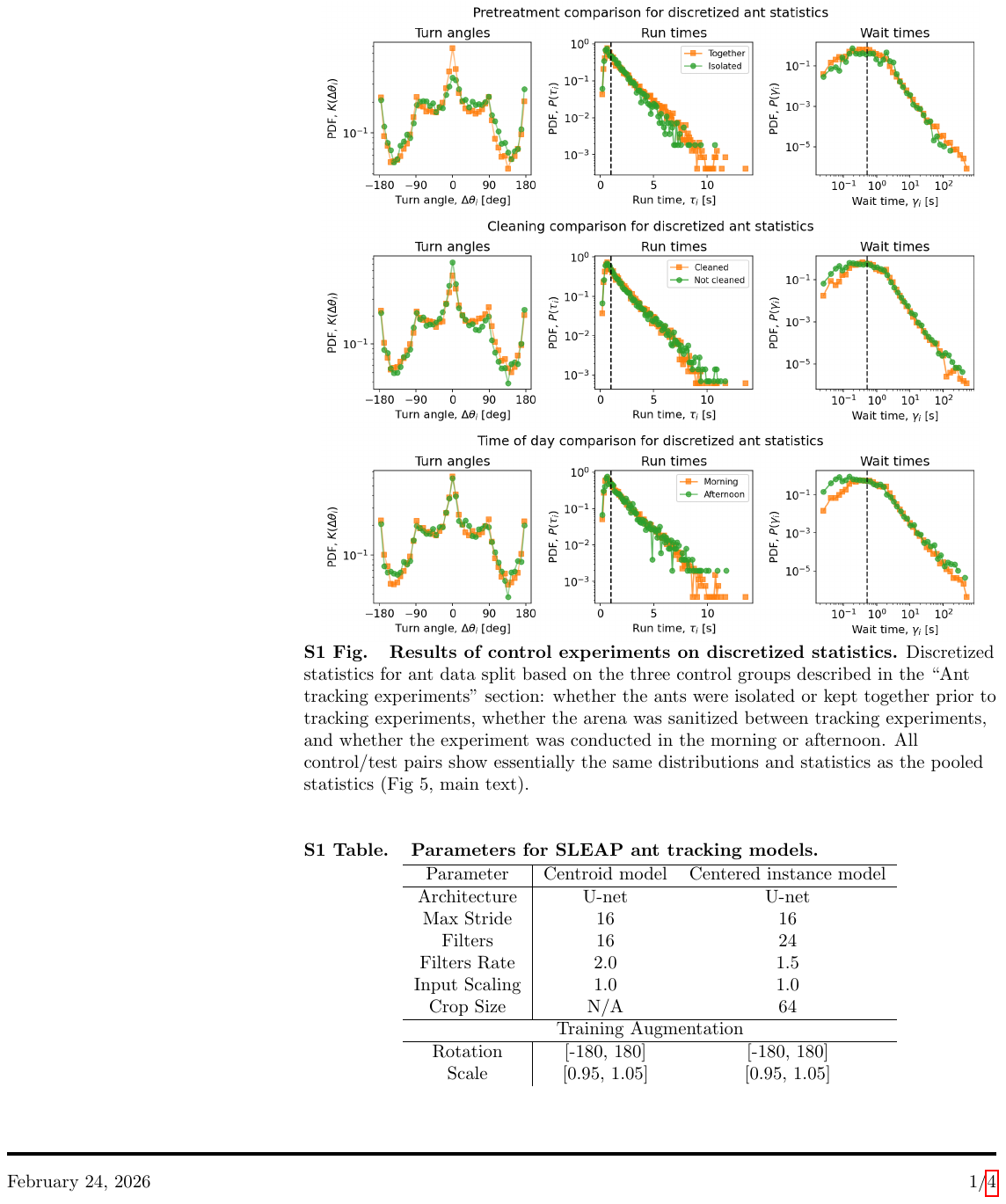}

\usepackage[right]{lineno}

\usepackage[nopatch=eqnum]{microtype}
\DisableLigatures[f]{encoding = *, family = * }

\usepackage[table]{xcolor}

\usepackage{array}

\newcolumntype{+}{!{\vrule width 2pt}}

\newlength\savedwidth

\raggedright
\usepackage[aboveskip=1pt,labelfont=bf,labelsep=period,justification=raggedright,singlelinecheck=off]{caption}

\makeatletter
\renewcommand{\@biblabel}[1]{\quad#1.}
\makeatother

\usepackage{lastpage,fancyhdr,graphicx}
\fancyheadoffset[L]{2.25in}
\fancyfootoffset[L]{2.25in}
\usepackage{longtable}

\usepackage{graphicx}
\graphicspath{{./images/}}

\usepackage{subcaption}

\begin{document}
\vspace*{0.2in}

\begin{flushleft}
{\Large
    \textbf{Stochastic modeling of long-legged ant \textit{A. gracilipes} locomotion in laboratory experiments} 
}
\newline
\\
Jack Featherstone\textsuperscript{1},
Anouk Béraud\textsuperscript{1},
Meta Virant-Doberlet\textsuperscript{2},
Antonio Celani\textsuperscript{3},
Mahesh Bandi\textsuperscript{1,*}
\\
\bigskip
\textbf{1} Nonlinear and Non-equilibrium Physics Unit, Okinawa Institute of Science and Technology, Onna, Okinawa, Japan, 904-0495
\\
\textbf{2} National Institute of Biology, Ljubljana, Slovenia, 1000
\\
\textbf{3} Abdus Salam International Centre for Theoretical Physics, Trieste, Italy, 34151
\\
\bigskip

* bandi@oist.jp

\end{flushleft}
%
\section*{Abstract}
Stochastic modeling of movement behavior provides a valuable way to understand how complex motion can be generated from relatively simple building blocks. Ants demonstrate sophisticated social behavior ranging from foraging to nest relocation; while emphasis is often placed on the communication methods used to synchronize individuals, the movement paradigms of those individuals are of tantamount importance. Here, we apply a stochastic modeling approach to better understand the movement of isolated long-legged ant (\textit{A. gracilipes}) specimens, informed by extensive laboratory tracking experiments. We find that a combination of active Brownian and run-and-tumble models reproduces the trajectory statistics observed in experiments, both qualitatively and quantitatively. We identify reproducible probability distributions for the turn angles, run times, and waiting times across specimens, and find good agreement between analytical predictions and quantities empirically measured from the trajectories. Having such a model allows for a better understanding and predictions of movement ecology from both simulations and analytics, and even can give insight into the underlying generative mechanisms of motion and the ants' sensory systems. 

%
\section*{Author summary}
Being able to identify a simple model of motion for a potentially very complex organism is a tantalizing goal, and has lead to the growth of the field of movement ecology over the past few decades. Stochastic modeling has been widely applied to animal motion as a tool to this end, though many applications end up with an incomplete view of such a process by only considering things like step-size distributions in isolation. Here, we apply stochastic modeling to experimental ant (\textit{A. gracilipes}) motion data to not only identify probability distributions of relevant quantities, but also to make comparisons to analytically-derived results. This gives a qualitative and quantitative understanding of the different phases that comprise motion, and offers a description of the individual behavior of ants. Further, the stochastic model can be leveraged to make predictions about the ants' behavior, as well as to simulate individual trajectories.



\section*{Introduction}

Movement is central to just about every facet of life for organisms spanning from the smallest bacteria to the largest mammals. The path an animal or other organism follows is affected by a wide range of internal and external factors including anatomy \cite{pontzer_EffectiveLimb_2007, watari_HydrodynamicsRunandTumble_2010}, sensory capabilities \cite{pratt_UseEdges_2001}, hunger \cite{wallin_InfluenceHunger_1994}, the presence of other individuals \cite{hu_EntangledActive_2016, tan_AntipredatorDefences_2024}, and environmental structure \cite{johnson_AnimalMovements_1992}. Quantifications of an organism's motion, typically involving analyses of trajectories or movement statistics, are crucial to understanding their ecology, biology, and behavior --- altogether referred to as their \textit{movement ecology} \cite{mendez_StochasticFoundations_2014}. This plays a role in how and why we find populations distributed across continents, though also how an individual explores its local surroundings.

Ants are particularly well-studied when it comes to movement and foraging behavior \cite{traniello1989foraging}, primarily because of their rich collective interactions \cite{lizonalallemand_SophisticatedModular_2010, feinerman_PhysicsCooperative_2018}. This  wealth of work includes experiments conducted in natural environments \cite{abbott_SupercoloniesInvasive_2005, win_SeasonalTemporal_2018} and in the laboratory \cite{dussutour_NoiseImproves_2009, chapman_BehaviouralSyndromes_2011}, and across a wide range of species. Indeed, these collective interactions are often so rich that the movement ecology of individuals has been understudied in comparison, leaving many questions about foraging and navigation on the single specimen scale unanswered. We explore the structure and statistics of motion on this individual specimen scale here, focusing on one particular species, the long-legged ant, \textit{Anoplolepis gracilipes} \cite{chong_InfluencesTemperature_2009}. This species is chosen not only for it's convenience and availability, but also it's international (invasive) presence, and the fact that it has been shown to demonstrate particularly efficient foraging and recruitment behavior \cite{lizonalallemand_SophisticatedModular_2010}. This efficient social behavior has led to these ants establishing overwhelming dominance in non-native areas \cite{abbott_SupercoloniesInvasive_2005}. While the communication and synchronization methods using pheromones play a large role in this dominance, the relatively microscopic behavior of individuals also plays an important role in building up this collective interaction.


Owed to both their invasive capabilities and ongoing urbanization of previously wooded areas, the contexts in which these ants are found is becoming more and more synthetic --- in what resources are available, what potential shelters are made from, and how open or confined their foraging grounds are \cite{rajesh_InteractiveEffects_2020}. Previous work has demonstrated that, even in confined spaces, various insects, including locusts  \cite{bazazi_IntermittentMotion_2012a} and \textit{Myrmica} ants \cite{chapman_BehaviouralSyndromes_2011} display rich behavior when it comes to locomotion. The biggest effect of such artificial confined spaces often has to do with the edges or boundaries. Thigmotactic or wall-following behavior can reveal important information about the behavioural responses of the organism \cite{sanmartin-villar_EarlySocial_2022, detrain_DifferentialResponses_2019}, and even about the capabilities of their sensory systems \cite{pratt_UseEdges_2001}.

Beyond a descriptive approach of an organism's motion, we are often interested in understanding how that motion arises, which stimuli influence it, and what interactions might arise in different environments. One particularly useful tool in this regard is stochastic modeling \cite{mendez_StochasticFoundations_2014}, which allows us to construct probabilistic models from relatively simple building blocks. These building blocks are synthesized from statistics observed in real locomotion data, for example that the step sizes or waiting times of an organism seem to follow a particular probability distribution \cite{kramer_BehavioralEcology_2001, bazazi_IntermittentMotion_2012a}.

Regardless of which particular building blocks we use, having a model that reflects the statistics of real motion data affords many advantages. In the best case, one can actually describe the internal neural mechanisms of an animal using the very same stochastic mechanisms that characterize the resulting trajectories \cite{mazzucato_NeuralMechanisms_2022}. Otherwise, modeling motion in this way allows us to explore how the organism might interact with some external environment, especially relating to searching or foraging behavior \cite{pezzotta_OptimalSearch_2018}. For example, one can predict how well a searcher/predator would fare given a specific distribution of rewards/prey \cite{viswanathan_OptimizingSuccess_1999}, and use that to extrapolate information about the ecology of the organism in a real environment \cite{sims_ScalingLaws_2008}. Similarly, many studies on chemotaxis and olfactory searching illuminate what strategies work well when navigating in (potentially weak) gradient fields \cite{pierce-shimomura_FundamentalRole_1999, celani_BacterialStrategies_2010, celani_OdorLandscapes_2014, kirkegaard_RoleTumbling_2018}. Finally, beyond understanding the statistical properties of motion, having a model of an organism's motion allows us to simulate individual trajectories. This is useful for a wide variety of purposes, with applications ranging all the way from simulating stimuli for real specimens \cite{chouinard-thuly_TechnicalConceptual_2017}, to creating video games \cite{tu_ArtificialAnimals_1999}, to planning roads \cite{mohamad_AntColony_2005}, to invasive species dispersal predictions \cite{carrasco_UnveilingHumanassisted_2010}.


Reference \cite{sakiyama_LevylikeMovements_2016} propose a simulation technique for ant motion, but it only attempts to capture one-dimensional step-size distributions. We are not aware of any work that has investigated the statistical properties of an individual specimen's motion beyond simple metrics like the mean-squared displacement. Previous work has proposed models for (non-ant) motion in confined spaces, though this typically makes use of separate modeling approaches near the walls and in the open bulk \cite{jeanson_ModelAnimal_2003}. In contrast, we look to capture the full range (in an information sparse environment) of motion by first modeling the unbounded motion, and then introducing appropriate boundary conditions that correct this motion near the wall. The advantage here is that this allows a more typical treatment of the problem using the tools of stochastic physics, since the posited model has well-defined (and somewhat common) governing equations.

We first describe the ant tracking experiments which form the foundation of our model development. We then describe the proposed model, including analytically-derived expressions that allow us to make predictions and gauge how well this model applies to the experimental trajectories. We then describe how exactly these experimental trajectories are compared to the model, and how statistical distributions are extracted from the sampled data. We share these statistical distributions, which show good agreement with those expected given the stochastic model. Finally, we demonstrate how this model allows for the simulation of ant-like trajectories, whose statistics recreate features observed in the experimental data.

\section*{Materials and methods}

\subsection*{Ant tracking experiments}

\begin{figure}
    \centering
    \includegraphics[width=0.8\textwidth]{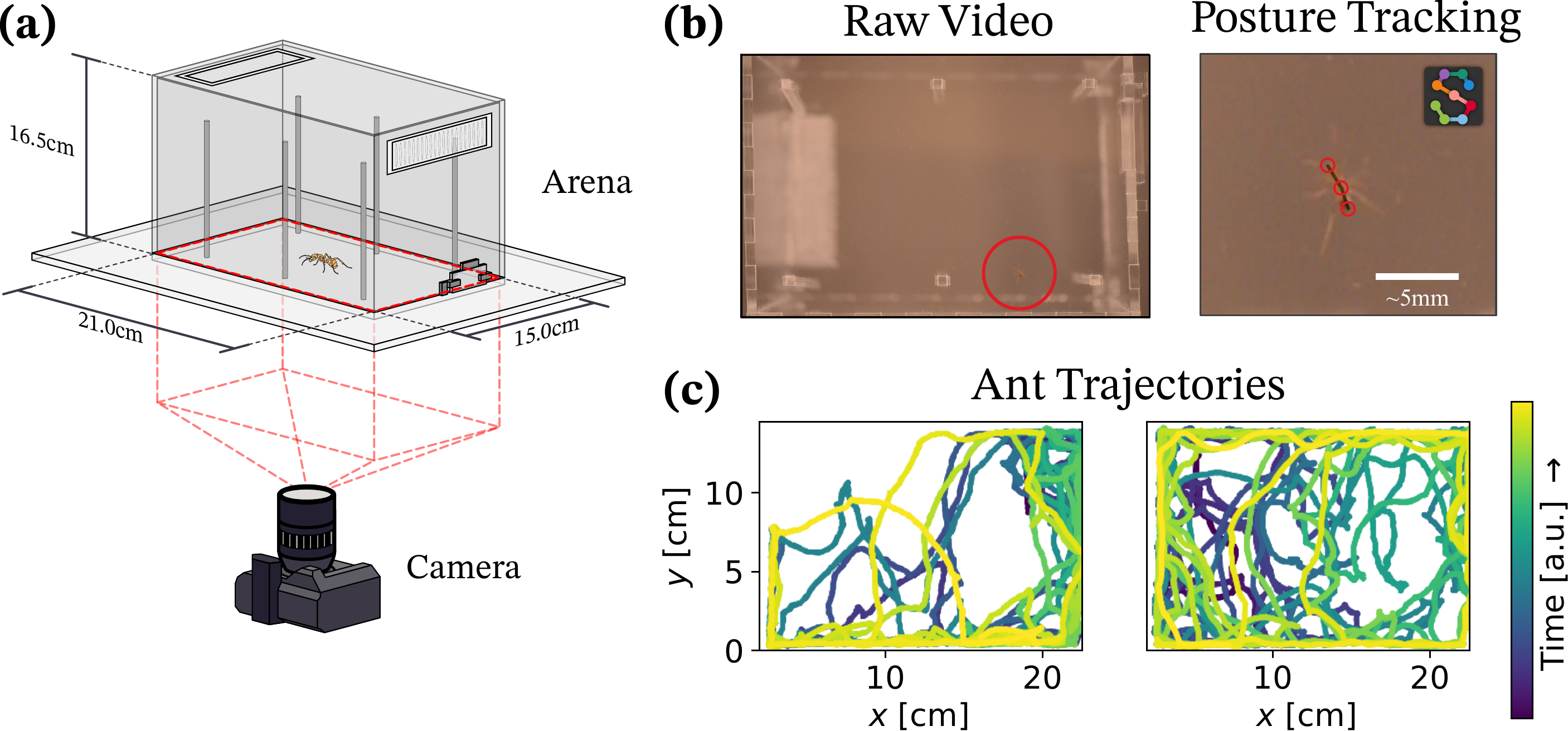}
    \caption{\textbf{Overview of experiment to extract ant trajectories in a confined arena.} Specimens are placed alone in an acrylic enclosure and recorded for 20 minutes. Ant position is tracked
    from videos using a neural network tracking approach (SLEAP), giving spatial trajectories.}
    \label{fig:overview}
\end{figure}

We performed ant tracking experiments using {\it Anoplolepis gracilipes} (``yellow crazy ant'' or ``long-legged ant'' \cite{lee_BiologyEcology_2022}) worker specimens (approx. $5$mm long) collected from around the campus of the Okinawa Institute of Science and Technology Graduate University, in Okinawa, Japan. After being collected manually, specimens were placed in a small container (approx. 5x5x10cm) potentially with conspecifics from the same area, for up to a few hours. One-by-one, specimens were placed alone in an acrylic, rectangular arena of size 15x21x16.5cm (Figure \ref{fig:overview}a), where they were recorded using a Nikon D800E camera at $60$ Hz for roughly 20 minutes. This arena size is typical of works interested in the confined behavior of insects on the scale of \textit{A. gracilipes} \cite{chapman_BehaviouralSyndromes_2011}. Experiments were performed over several weeks between May 2024 and July 2024. Each ant was used only once.

Control trials were performed to confirm that being placed with conspecifics prior to the tracking experiments did not have a significant effect on ant's motion. Ants were either isolated after collection, or placed with conspecifics, but in all cases the tracking experiments were always performed with only a single ant present.

Control trials were also performed in which the enclosure was sanitized --- washed with soap and water, then cleaned with 70\% ethanol \cite{jeanson_ModelAnimal_2003} --- between each experiment, or left as is. Yellow crazy ants are known to produce several types of pheromones for navigation and orientation \cite{lee_BiologyEcology_2022, lizonalallemand_SophisticatedModular_2010}, so these trials were intended to determine if there were stimuli induced from previous ants in the otherwise static environment.

Finally, all trials were performed during either the morning (08:00 - 12:00) or afternoon (12:00 - 17:00). It has been suggested that these ants tend to be more active in the morning \cite{chong_InfluencesTemperature_2009}, so we looked to compare if the time of day of the experiment affected the statistics.

We saw no significant differences in any of the trajectory statistics presented in this work across all of these control-test pairs (\nameref{fig:S1}, compare to Fig \ref{fig:discretized_stats}). This is not to say that these factors had no effect at all on the trajectories, but just that the quantities of interest were robust to these changes in experimental protocol. As such, the figures in the main text of this work include data from both the control and test groups for the above three comparisons.

A limited set of experiments was performed with a different camera (Phantom v641 high-speed camera) that is capable of shooting at a much higher framerate and resolution ($120$ and $300$ Hz, 2560x1600 pixels), in order to characterize the scale of temporal and spatial discretization artifacts in the trajectories. This dataset generally shows similar statistics to the main dataset, suggesting that the principal statistical properties are not discretization artifacts. While one can intuitively predict the scale of such artifacts {\it a priori} in other cases, the neural network tracking approach (and its uncertainty) makes this more difficult. No analysis of this dataset is included here, though the data is available in the online repository.

Other factors of collection and treatment were kept constant where possible. Temperature and humidity were kept roughly constant around $22$\textdegree C and $50\%$ relative humidity by the climate control in the laboratory for the duration of experiments. This is towards the lower end of, though still within, the temperature range for which these ants forage \cite{haines_ImpactControl_1994, chong_InfluencesTemperature_2009, lee_BiologyEcology_2022}. Overhead fluorescent lights were kept on for all experiments, though lighting is suspected not to affect these ants' activity \cite{chong_InfluencesTemperature_2009}.

\subsection*{Trajectory analysis}

Following experiments, the video data was consolidated and used to train a deep convolutional neural network to track the posture of the ants using SLEAP \cite{Pereira2022sleap}. A multi-animal (top down) model is used, composed of two U-net neural networks \cite{ronneberger_UNetConvolutional_2015}, one which identifies the rough location of the ant (``centroid model''), and the second which identifies specific keypoints in a limited region around the identified location (``centered instance''). This latter network tracks three keypoints on the ants: the tip of the head, the center of the thorax, and the rear tip of the abdomen. This structure is preferred to a single-animal model since the ant is so small compared to overall arena size. The parameters of these networks are described in \nameref{tab:S1}.

The positions of the keypoints are extracted to give the spatial trajectory of each body part through time. After training on 280 manually annotated frames, we observe an average error per keypoint of $\sim 0.2$ mm, which is estimated to be of similar magnitude to the error in the annotations of these features. All keypoints within a single frame are averaged to find the center-of-mass position of the ant; we don't consider the relative motion of the ant's body in this work. To further reduce the effect of artifacts, we perform windowed averaging (smoothing) on the center-of-mass position of the ants with a window size of $10$ (representing a time window of about $0.15$ seconds, chosen empirically).

This tracking model was used to extract the trajectories of all ants across all videos. In some cases, the trajectory is split into separate parts because either the ant climbs the walls of the arena --- and therefore leaves the camera view --- or the tracking algorithm fails. The latter case mostly consists of very short interruptions, which can be fixed by linearly interpolating the trajectory for the missing points. We perform this interpolation up to a maximum of $30$ missing data points (chosen empirically, $0.5$ seconds in real-time). If there is a gap larger than $30$ frames, we don't interpolate the trajectory as it could introduce artifacts.
Most trials had a few long segments representing the majority of the motion, with some shorter segments punctuating these longer ones. Except for computing the occupation maps, these short segments are excluded from analysis by taking only segments longer than a cutoff time, $t_{\text{min}} = 5$ seconds (chosen empirically).

We mark 11 trials as outliers due to various experimental differences to the rest of the data, which are summarized in \nameref{tab:S2}. Excluding these trials results in an ensemble of ant segments representing more than 24 hours of footage capturing the motion of nearly 100 individual specimens. Some example segments are shown in Fig \ref{fig:overview}c.

\subsection*{Ant motion as a run-and-tumble process}

We propose that the run-and-tumble \cite{fodor_StatisticalPhysics_2018, fu_FokkerPlank_2021, zamora_ExploringRunandtumble_2024} framework qualitatively matches the typical movement patterns of these ants. This stochastic process involves short time-scale ballistic motion, punctuated by discrete reorientation events. Given a particle (or ant) with position $\vec x$ and orientation $\theta$, the most basic Langevin equation for this process in two dimensions looks like:
\begin{align*}
    \frac{d}{dt} \vec x = &~ v \hat e(t)\\
    \frac{d}{dt} \theta = &~ \sum_i \Delta \theta_i \delta (t - t_i).
\end{align*}
The first equation represents the translational motion of the particle, implementing the ``run'' evolution, with speed $v$ (taken as a constant) and direction vector $\hat e$:
\begin{equation*}
    \hat e(t) = \cos(\theta(t)) \hat x + \sin(\theta(t)) \hat y.
\end{equation*}

The second equation describes the tumbling behavior of the particle, where the orientation is updated at discrete times $t_i$ by some amount $\Delta \theta_i$. In principle these could be drawn from any arbitrary distribution, though common choices for these two are:
\begin{equation}
    \label{eq:angle_kernel}
    \Delta \theta_i \leftarrow K(\Delta \theta_i) = \frac{1}{2 \pi I_0(\sigma)} e^{\sigma \cos(\Delta \theta_i)},
\end{equation}
\begin{equation}
    \label{eq:tumbling_times}
    t_{i+1} - t_i = \tau_i \leftarrow P(\tau_i) = \frac{1}{\tau} e^{-\tau_i / \tau}.
\end{equation}
That is, the reorientation kernel $K$ is a zero-centered von Mises distribution with width parameter $\sigma$ ($I_0$ is the modified bessel function of the first kind of order zero), and the time between tumbles (run time) is drawn from an exponential distribution with average $\tau$. This is equivalent to defining a constant tumbling rate $\frac{1}{\tau}$. As we will see,  these choices are well suited for the ant data. On top of this basic model, we consider three additions to more realistically model the ant motion: boundary conditions, rotational diffusion, and finite tumbling times.

Other studies have proposed piecewise models that evolve according to different governing equations in the bulk and near the walls \cite{jeanson_ModelAnimal_2003}. Here, we propose to model our data with a more unified approach: using the same fundamental dynamics across the entire space, though with a local boundary force/torque at the edge of the arena. The fundamental difference is that the piecewise approach introduces non-local effects of the boundaries, invoking some assumptions about the sensory system of the specimen (ie. how does the insect ``know'' that it is at the boundary?). Such assumptions can be very helpful, though our approach provides a more minimal (and therefore, hopefully more widely applicable) model, which might be expanded based on the anatomy and behavior of a specific organism.

Including rotational diffusion allows the model to capture some of the roughness and curvature that real ants demonstrate when walking. This introduces a qualitative separation between small angle changes that result in curvature along an otherwise straight path, and large angle changes that constitute reorientation behavior \cite{bartumeus_FractalReorientation_2008}.

Finally, we introduce waiting times at each tumble, a phenomenon that is repeatedly observed both in our data and in that for other insects \cite{bazazi_IntermittentMotion_2012a}. This is implemented via a ``waiting function`` $H(t)$ that is zero during waiting times of length $\gamma_i$, and 1 otherwise,
\begin{equation}
        H(t) = 1 - \sum_i \sigma(t - t_i) - \sigma(t - t_i- \gamma_i) \label{eq:lang_H},
\end{equation}
where $\sigma(t)$ is the Heaviside function. For now, we make no assumptions about the form of the distribution from which the $\gamma_i$ are drawn.

Altogether, we have the governing equations:
\begin{align}
    \frac{d}{dt} \vec x = &~ v \hat e(t) H(t) + \vec F_x \label{eq:lang_x}\\
    \frac{d}{dt} \theta = &~ \sum_i \Delta \theta_i \delta (t - t_i) + \sqrt{2 D_r} \xi(t) H(t) + F_r \label{eq:lang_theta}.
\end{align}
The translational boundary force $\vec F_x$ is taken to be simply a rigid body force along the boundary normal, $\hat n$:
\begin{equation}
    \vec F_x = -v (\hat e(t) \cdot \hat n) \hat n.
\end{equation}
This is equivalent to defining an infinite well potential that the process evolves within. The rotational boundary condition is taken to be a torque that aligns the particle with the closer tangent of the wall:
\begin{equation}
    \label{eq:torque_boundary}
    F_r = \text{arccos} (\hat e(t) \cdot \hat n) ~ \text{sign}(\hat e(t) \cdot \hat n).
\end{equation}
Finally, $\xi(t)$ is a delta-correlated random variable with unit variance.

Simulations are most easily implemented using the Langevin equations (Eqs \ref{eq:lang_x}, \ref{eq:lang_theta}), though we can also formulate a Fokker-Planck equation for the probability density \cite{celani_BacterialStrategies_2010, kirkegaard_RoleTumbling_2018}. This is particularly useful in deriving the expectation values of various quantities, for example, the persistence time (full derivation in \nameref{appendix:S1}),
\begin{equation}
    \label{eq:persistence}
    \tau_p = \frac{1}{D_r + \tau^{-1} \left( 1 - \frac{I_1(\sigma)}{I_0(\sigma)} \right)},
\end{equation}
where $I_1$ is the modified bessel function of the first kind of order one. This does not take the finite size of the arena into account, and thus is only valid if we consider the motion in the center of the arena. The typical way to compute the persistence time of an arbitrary trajectory (ie. imposing no model assumptions), which we compare to the above expression, is to consider the autocorrelation of the orientation through time. For a stationary, persistent process, we have:
\begin{equation}
    \label{eq:cos_persistence}
    \langle \cos\theta(t) \cos\theta(t + \Delta t) \rangle \sim \langle \cos\theta(\Delta t) \rangle
    \sim \text{exp}(-\Delta t / \tau_p).
\end{equation}
By fitting this function with an exponential form, we can extract the persistence time $\tau_p$. In subsequent sections, this is referred to as the ``direct fit'' method, whereas the formula derived from the discretized model is referred to as the ``discrete fit'' method.

We can make further use of this Fokker-Planck formulation by considering the run-and-tumble model as an active Brownian process with stochastic resetting \cite{gupta_StochasticResetting_2022}. This allows us to calculate the typical velocity from the short timescale mean-squared displacement (see \nameref{appendix:S1} for more details),
\begin{equation}
    \label{eq:msd_v}
    \langle r(\Delta t)^2 \rangle \approx v^2 \Delta t^2 \hspace{1cm} (\Delta t < \tau),
\end{equation}
as well as the rotational diffusion coefficient from the persistence time \textit{during runs},
\begin{equation}
    \langle \cos\theta(\Delta t) \rangle
    \sim \text{exp}(-D_r \Delta t) \hspace{1cm} (\Delta t < \tau).
\end{equation}
As above, computing the velocity by fitting the MSD is referred to as the ``discrete fit'' (since it invokes the model), whereas computing the median velocity from the derivative of the trajectory is referred to as the ``direct fit''.

\subsection*{Trajectory discretization}

\begin{figure}
    \centering
    \includegraphics[width=\textwidth]{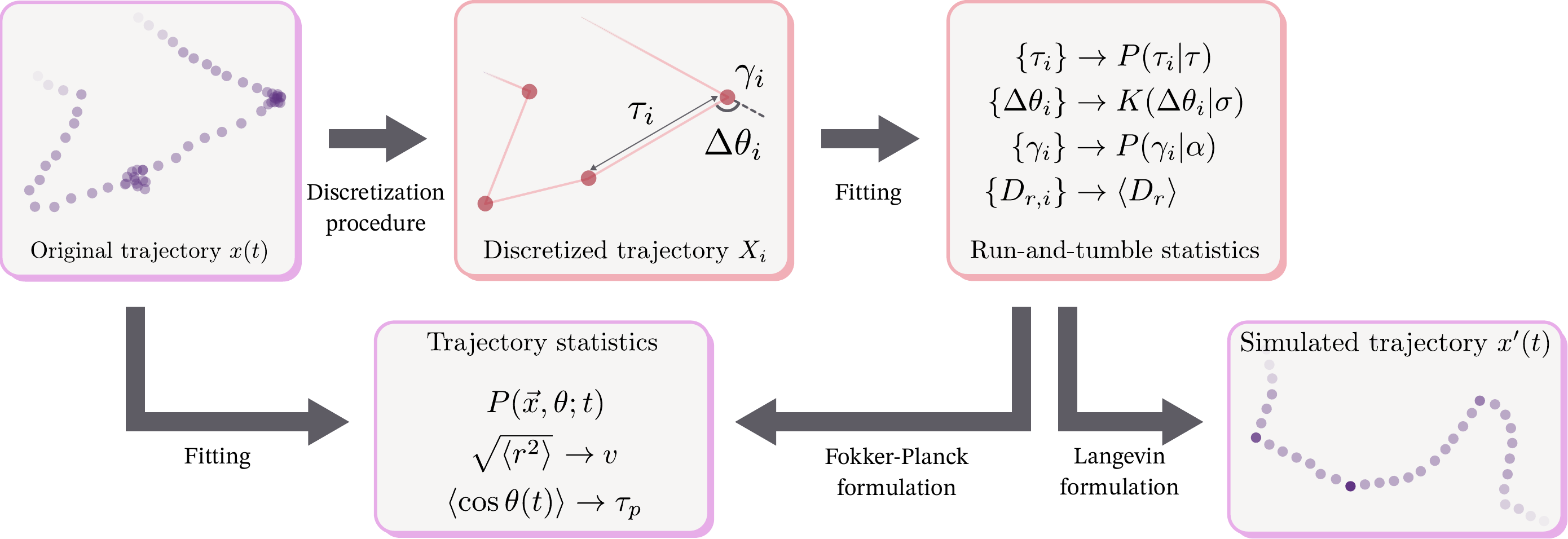}
\caption{\textbf{Trajectory analysis workflow.} Original sampled trajectories are used to compute trajectory statistics like occupation maps and mean squared displacement, as well as to identify points when the ant changes direction and/or stops moving, generating a \textit{discretized} trajectory. This allows for the extraction of the times spent moving in a straight line, $\tau_i$, the angle changes between these segments, $\Delta \theta_i$, the time spent waiting before each reorientation, $\gamma_i$, and the angle drift during each straight run, $D_{r,i}$. These values are used to derive statistics about the original trajectory using a Fokker-Planck equation, and to simulate new ant-like trajectories using a Langevin equation.}
    \label{fig:discretization}
\end{figure}

In order to compare the experimental data with the stochastic model described in the previous section, we have to identify discrete run events, as well as the tumbling events that punctuate them (Fig \ref{fig:discretization}) in this trajectory data. The idea of discretizing a continuous trajectory has been both applied and discussed at length in the movement ecology literature \cite{bartumeus_AnimalSearch_2005, reynolds_DisplacedHoney_2007, kawai_MultiscaleProperties_2012, mendez_StochasticFoundations_2014}. Such an analysis can reveal important information about the structure of movement \cite{edelhoff_PathSegmentation_2016}, or the typical transitions made while moving \cite{junot_RuntoTumbleVariability_2022}. The technical application of this method is usually relatively simple, and involves breaking up the trajectory at points when the animal turns more than some critical angle, $\theta_c$ and/or the speed is less than some threshold, $v_{c}$. That being said, the justification for selection of parameter values for this technique is often not straightforward, and, in particular, the inherent subjectivity of this analysis is often criticized \cite{pyke_UnderstandingMovements_2015}: it is often the case that many different conclusions can be drawn from the same dataset by choosing different discretization parameter(s), all of which might seem reasonable at first glance.

This isn't an issue when one can choose the discretization parameters \textit{a priori} based on properties of the system being investigated, or when there is no ambiguity as to the different phases of behavior. For example when the behavior of the animal is strongly intermittent and/or when the motion is considered in only one dimension, it becomes very obvious when the organism is ``running'' and when the animal is ``tumbling'' \cite{buhl_DisorderOrder_2006, bazazi_IntermittentMotion_2012a}. Unfortunately, most animal behavior doesn't demonstrate such a dichotomy, and we often have to determine the scale at which we discretize \textit{a posteriori}. In this case, we may still apply this technique to a trajectory, though we should confirm several pieces of information to minimize the amount of subjectivity in the results. First, that it is reasonable to assume that there even exists a scale at which discretizing the trajectory gives relevant information about the underlying continuous process. In other words, does it, both qualitatively and quantitatively, make sense to model this process as having discrete phases? Second, that the discretization process is able to reliably extract the desired statistics from trajectories with known properties. And third, that the results are not especially sensitive to the choice of discretization parameter(s).

We use an accumulator discretization method (sometimes called ``non-local'' \cite{reynolds_DisplacedHoney_2007}) in which the moving average of the current run direction is computed, and when the angle between that average direction and the next position is greater than some threshold value $\theta_c$, a new run is defined. We also segment the trajectory by times when the velocity falls below a threshold $v_c$, even if the ant continues moving in the same direction after resting. This combination allows for detection of runs punctuated either by changes in direction or waiting times (Fig \ref{fig:discretization}). From this discretized trajectory, we extract the distribution of turn angles $\Delta \theta_i$, run times $\tau_i$, and waiting times $\gamma_i$. As discussed in the previous section, we also compute the rotational diffusion coefficient $D_r$ and velocity $v$ using expressions derived from the model in the short time limit. All of these analyses are tested on synthetic run-and-tumble data generated using Eqs \ref{eq:lang_x}, \ref{eq:lang_theta}, shown in \nameref{fig:S1}; in general, the technique does a reasonable job of estimating the underlying parameters, including for data generated in the exact geometry used in the experiments.

All fitting is performed with either the \texttt{powerlaw} \cite{alstott_PowerlawPython_2014} or \texttt{scipy} \cite{2020SciPy-NMeth} packages in Python. When a centrality measure is needed to characterize statistics below, the median is preferred over the mean, since some of the analyses that involve curve fitting can yield unrealistic values when there are few data, which heavily skew the mean.

\section*{Results}
\label{sec:results}

\subsection*{Locomotion statistics}

\begin{figure}
    \centering
    \begin{subfigure}{0.46\textwidth}
    \begin{center}    
    \includegraphics[width=0.82\linewidth]{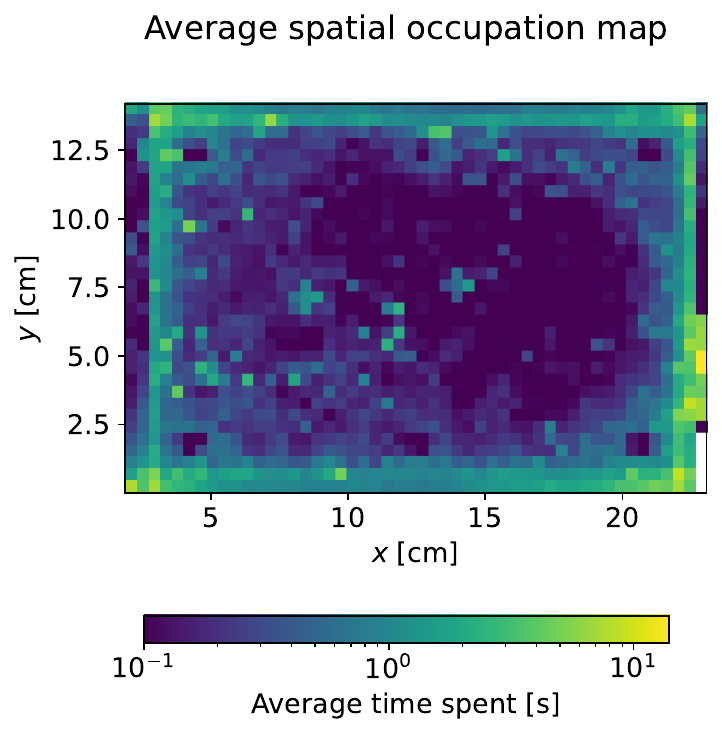}
    \end{center}
    \end{subfigure} 
    \begin{subfigure}{0.53\textwidth}
    \begin{center}    
    \includegraphics[width=0.9\linewidth]{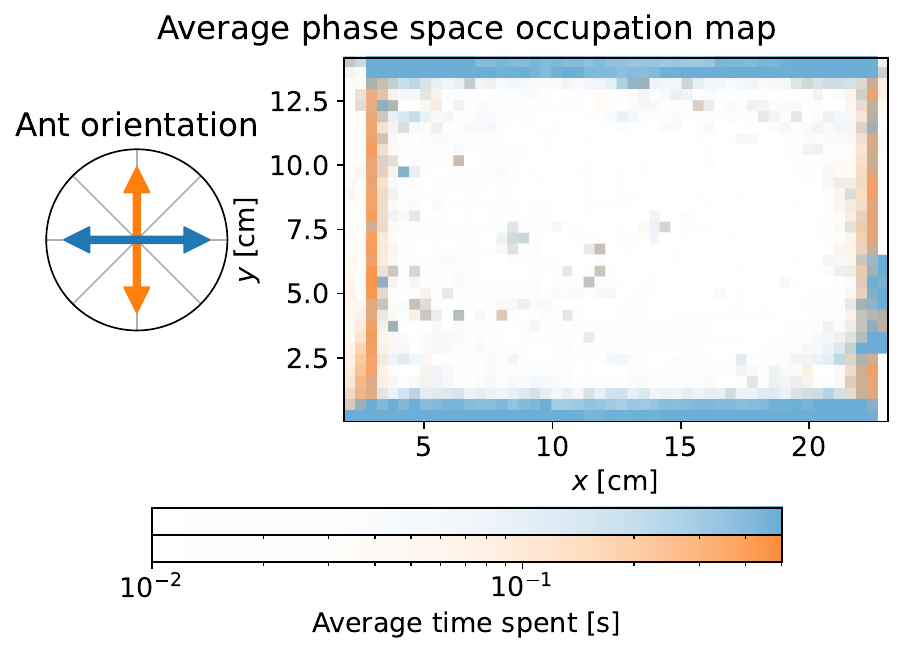}
    \end{center}
    \end{subfigure}

    \caption{\textbf{Spatial and phase space occupation maps for ant motion.} Average amount of time spent in each region of the experimental arena (left) and average amount of time spent in each region with a particular orientation (right). For phase space map, orange coloring represents time spent moving \textit{either} up or down, and blue represents moving \textit{either} left or right.}
    \label{fig:occupation_map}
\end{figure}

One of the most prominent features of the locomotion data, as one might expect, is the impact of the finite arena and its geometry. Many different types of ants are known to demonstrate thigmotactic or other wall-following behavior both in natural \cite{collett_SelectionUse_1997} and laboratory settings \cite{pratt_UseEdges_2001, sanmartin-villar_EarlySocial_2022, detrain_DifferentialResponses_2019}. We observe that the most densely visited locations are those on the boundaries, visualized in Fig \ref{fig:occupation_map}, left, through the average spatial occupation map. We can also observe how the specimens align their motion with the boundary by examining various cuts of the phase space (spatial position and orientation) map, shown in Fig \ref{fig:occupation_map}, right. When near the left and right (speaking from a birds-eye view, as in the figure) walls, the ants overwhelmingly move up or down, and vice-versa for the top and bottom walls.  

Beyond this stereotyped behavior around the boundaries, we are next interested in the statistics of motion in the center of the arena. Based on these occupation maps, we define the center of the arena by cropping data within $1.5$cm of the walls.

\begin{figure}
    \centering
    \includegraphics[width=0.6\textwidth]{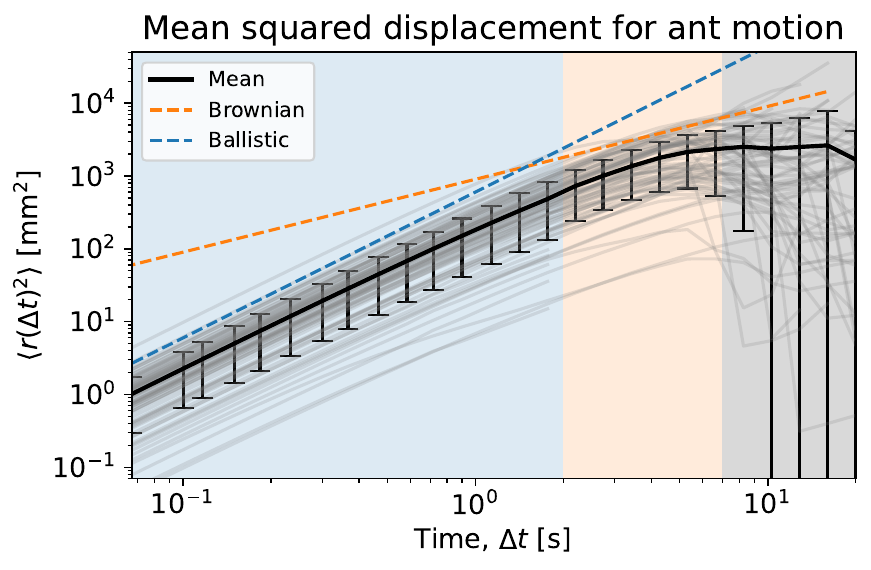}
    \caption{\textbf{Mean squared displacement (MSD) for ant motion.} Blue, orange, and grey shading indicates regions where the displacement scales linearly with time (ballistic motion), with the square root of time (Brownian motion), and independently of time (finite-size effects). Regions are drawn as rough guides, not based on specific quantifications. Faint curves are the MSD for each individual trial; solid black line represents the pooled statistics for all trials, with standard deviations shown as error bars.}
    \label{fig:msd}
\end{figure}

One of the most classic objects of interest in trajectory analysis is the mean-squared displacement (MSD), which is often used to identify whether a process is sub- or super-diffusive. Applied to the ant trajectories, this metric identifies three distinct regimes of motion, shown in Fig \ref{fig:msd}. At short time scales (blue shaded region), the ant moves ballistically, such that the displacement scales linearly with time, $\langle r^2 \rangle \sim t^2$, and the trajectory consists primarily of straight segments. At longer time scales (orange shaded region), we see roughly Brownian scaling of the MSD, $\langle r^2 \rangle \sim t$, indicating that the ant typically changes its orientation after the short time scale ballistic motion. Finally (gray shaded region), the finite size of the arena limits the distance the ant can displace from its previous positions, giving an MSD that no longer scales with time.

The location of the transition point between the ballistic and Brownian regimes is a (rough) quantification of the persistence time of the ant motion, ie. the typical time after which the ant reorients herself. As we will see, indeed the typical persistence time computed from both the direct and discrete methods below is near the transition point, $\sim 1$s. This transition point (and thus the persistence time) varies from experiment to experiment, which can be seen by looking at the MSD curves for each individual trial (faint curves in Fig \ref{fig:msd}). This clear separation in straight running behavior and reorientation behavior also provides quantitative evidence that indeed this process is well-described by a run-and-tumble model, justifying the application of the discretization procedure.

\subsection*{Discretized statistics}

We now apply the previously-described discretization procedure to the trajectory data, allowing us to better understand the contribution of different elements to the persistence. As previously discussed, the choice of parameters for this discretization process, $\theta_c$ and $v_c$, is important. That being said, we find that nearly all reasonable values of these parameters lead to the same trends in our data, with only minor quantitative differences (see \nameref{fig:S3}); as such, we show each result for only a single, representative, choice of discretization parameters, though we mention the range of typical values resulting from other parametrizations.

\begin{figure}
    \centering
    \includegraphics[width=\linewidth]{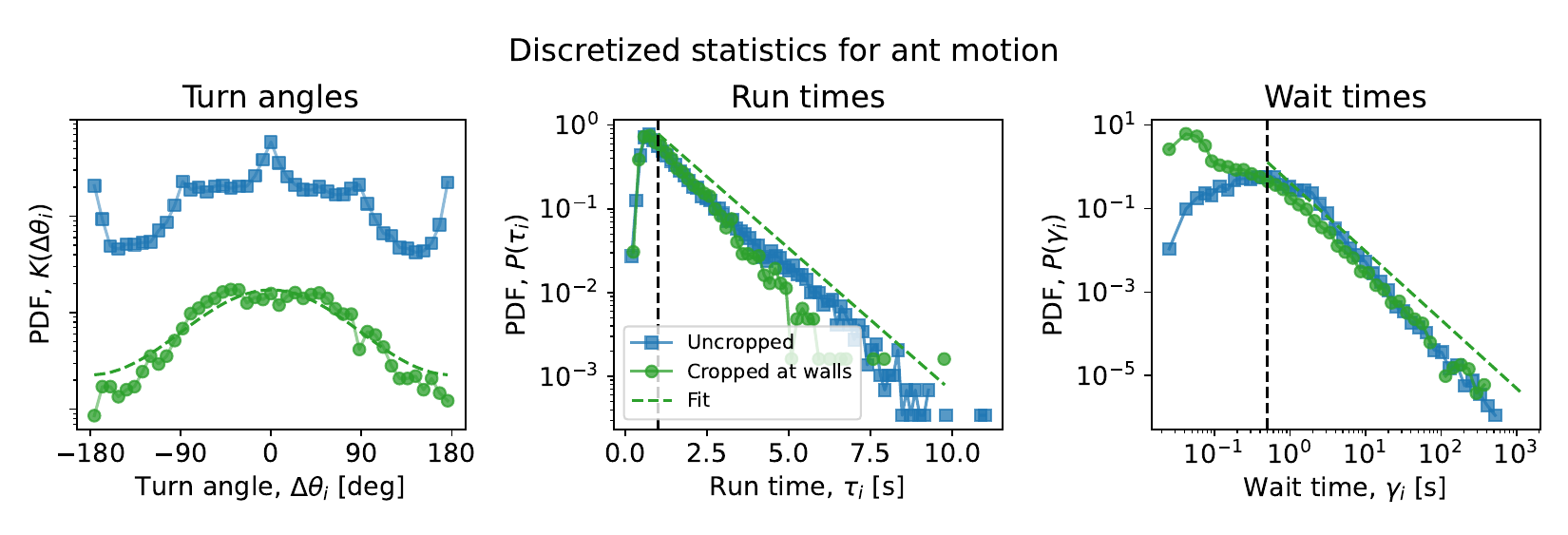}

    \caption{\textbf{Turn angle, run time, and wait time distributions for discretized ant motion.} PDFs are created using pooled data from all ant trials. A single, representative choice of discretization parameters ($\theta_c = 45$\textdegree, $v_c = 5$mm/s) is shown here; for comparison with other parameter choices, see \nameref{fig:S3}. Green dashed lines represent the median fit --- von Mises, exponential, and power law, respectively --- across the individual ant trials (in contrast to the pooled data shown in the curves with markers). Dashed vertical lines indicate the minimum value including in fitting.}
    \label{fig:discretized_stats}
\end{figure}

Shown in Fig \ref{fig:discretized_stats} are the turn angle, run time, and wait time distributions for the discretized trajectories.  We find that the ant shows very stereotyped behavior around the edges and corners of the arena, leading to a ``crown''-shaped turn angle distribution (Fig \ref{fig:discretized_stats}, left, blue squares) with peaks around $0$\textdegree, $\pm 90$\textdegree, and $\pm 180$\textdegree. When we consider only the central part of the arena, this turn angle distribution reduces to a unimodal distribution (Fig \ref{fig:discretized_stats}, left, green circles and dashed line). We fit this latter distribution with a zero-centered von Mises functional form, giving a width parameter $\sigma$ in the range $[0.6, 1.3]$ ($\sigma = 1.01$ for shown parameterization). Fixing the center of the distribution at zero assumes that the turn angle distribution is symmetric; this seems to be well justified, as allowing the center to vary as a fit parameter gives similar results, though with more noise.

The run time distribution has an exponential form for both the uncropped and cropped data, with decay constant (average run time) in the range $[1, 2.1]$ seconds  ($\tau = 1.27$s for shown parametrization). The wait time distribution has a power law form for both the uncropped and cropped data with an exponent in the range $[1.5, 1.7]$ ($\alpha = 1.64$ for shown parametrization).

\begin{figure}
    \centering
    \centering
    \begin{subfigure}{0.49\textwidth}
    \begin{center}    
    \includegraphics[width=0.85\linewidth]{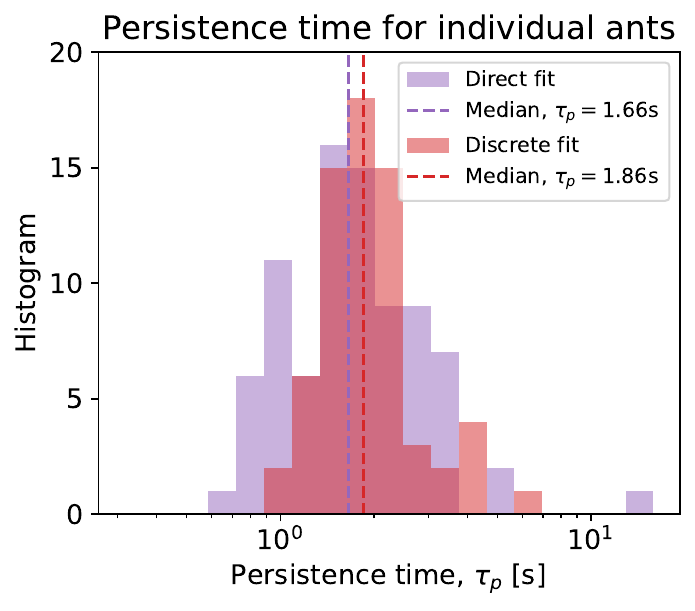}
    \end{center}
    \end{subfigure} 
    \begin{subfigure}{0.49\textwidth}
    \begin{center}    
    \includegraphics[width=0.87\linewidth]{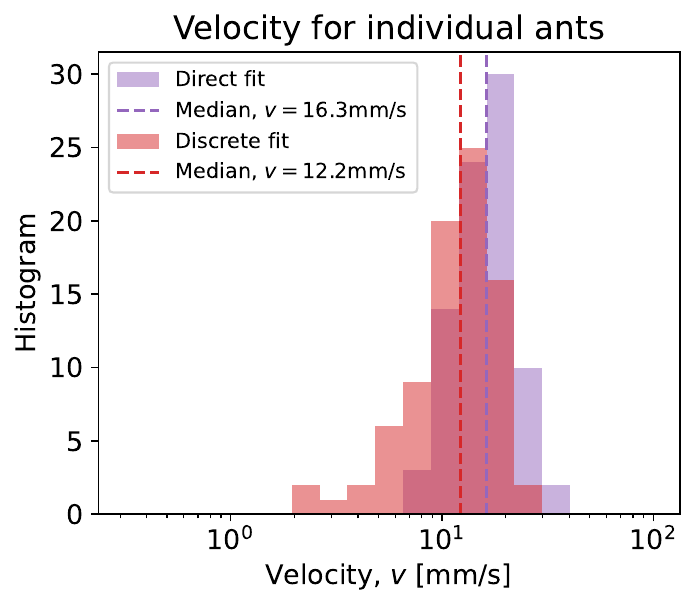}
    \end{center}
    \end{subfigure}

    \caption{\textbf{Persistence time and velocity for individual ants.} ``Direct fit'' distributions represent calculations of the persistence time and velocity without any model assumptions, while ``discrete fit'' distributions use an analytic expression derived assuming the proposed model.}
    \label{fig:comparisons}
\end{figure}

We estimate the value of the diffusion coefficient by fitting the persistence time \textit{within} each discretized run, as discussed previously. This, together with the values for $\tau$ and $\sigma$, allows us to calculate the predicted  persistence length of the process using Eq \ref{eq:persistence}. This is shown for each individual ant trial in Fig \ref{fig:comparisons}, labeled as ``discrete fit''. We can compare this to the persistence time directly calculated from the trajectories using Eq \ref{eq:cos_persistence}, also shown in Fig \ref{fig:comparisons} (``direct fit''). Both methods give similar distributions, with median values of $\tau_p = 1.86$s for the discrete fit and $\tau_p = 1.66$s for the direct fit.

We do a similar comparison for the typical velocity, computing this value either through expressions derived from the model (from Eq \ref{eq:msd_v}, ``discrete fit'') or by taking the derivative of the trajectory (``direct fit''), shown in Fig \ref{fig:comparisons}, right. As with the persistence, these two independent techniques gives very similar results, demonstrating the applicability of the proposed model.

\begin{figure}
    \centering
    \begin{subfigure}{0.49\textwidth}
    \begin{center}    
    \includegraphics[width=0.95\linewidth]{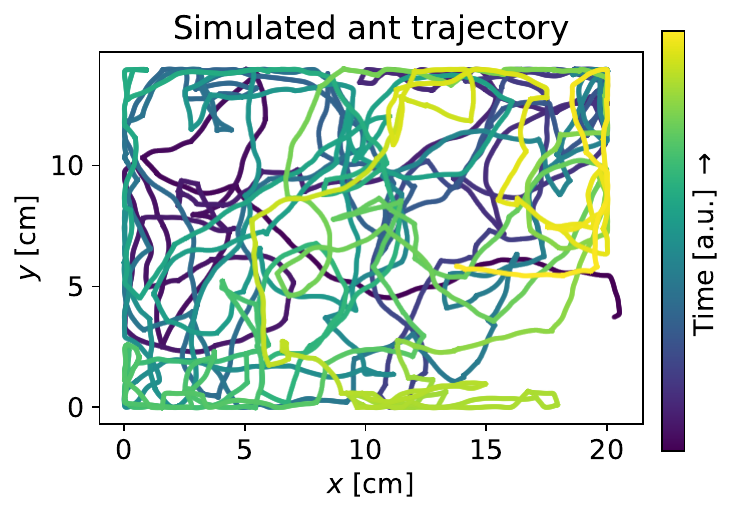}
    \end{center}
    \end{subfigure} 
    \begin{subfigure}{0.49\textwidth}
    \begin{center}    
    \includegraphics[width=0.95\linewidth]{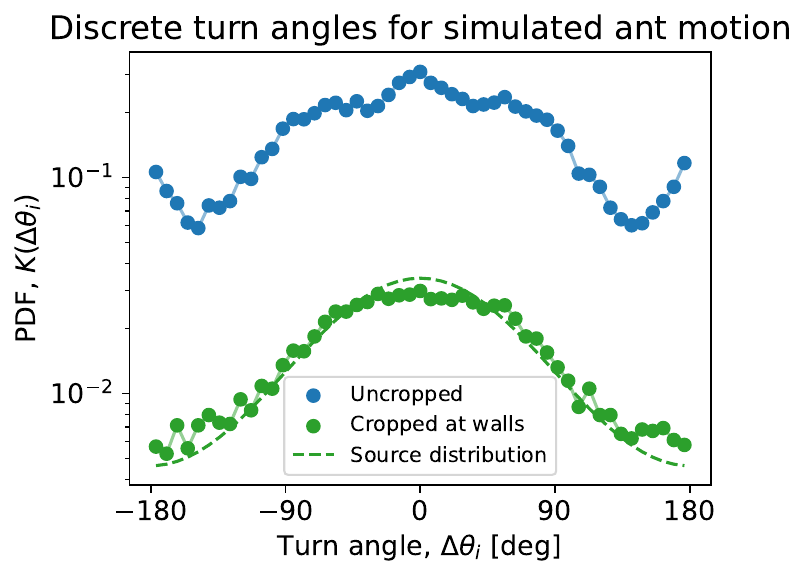}
    \end{center}
    \end{subfigure}

    \caption{\textbf{Simulated ant trajectories and their statistics.} A simulated trajectory using the Langevin equation (Eqs \ref{eq:lang_x}, \ref{eq:lang_theta}) with parameter values based on fits from experimental ant trajectories (left). The turn angle distribution for this simulated data qualitatively recreates the crown-shaped distribution from experimental trajectories (right).}
    \label{fig:simulation}
\end{figure}

Beyond guiding our analysis of the experimental trajectory data, one of the principle advantages of proposing a mathematical model of motion is that it allows us to simulate trajectories, as shown in Fig \ref{fig:simulation}, left. These simulated trajectories closely reproduce the statistics of the experimental data. For example, even though the simulation is derived from statistics in the center of the arena, the relatively simple boundary conditions implemented in the model qualitatively reproduce the multi-modal turn angle distribution throughout the entire arena (Fig \ref{fig:simulation}, right).

\section*{Discussion}

Interestingly, our distribution of wait times for \textit{A. gracilipes} is very similar to the wait time distribution for locusts in similar experiments \cite{bazazi_IntermittentMotion_2012a}. Both experiments yield power law distributions in the domain $10^{-1}s - 10^{2}s$, with exponents in the range $1.5 - 1.7$ (ours) and $1.67$ (theirs). This type of intermittent behavior has been suggested to indicate a searching paradigm where the agent is generally less reactive/perceptive during motion, and more so during pauses and rests \cite{benichou_TwodimensionalIntermittent_2006, benichou_IntermittentSearch_2011}. If we are to take these wait times as opportunities for the ant to sense and think about its surroundings, the power law form of the wait time distribution could suggest a (positive) relationship between the time invested and the volume of information in the sensory response. If the sensory response were relatively independent of the time invested, we would instead expect a wait time distribution with some typical scale, eg. an exponential or Gaussian. On the other hand, it is also possible that these pauses originate because of other considerations, for example, that the ant is conserving or recovering energy.

We don't see any evidence that the movement statistics of \textit{A. gracilipes} are scale-free, ie. that the run time distribution has a power law form, $P(\tau_i) \sim \tau_i^{-\beta}$. Such a form has been proposed principally in connection to the Levy Foraging Hypothesis \cite{viswanathan_OptimizingSuccess_1999}, for which many studies have expressed support \cite{sims_ScalingLaws_2008, sakiyama_LevylikeMovements_2016} or dissent \cite{edwards_RevisitingLevy_2007, sims_MinimizingErrors_2007, benichou_IntermittentSearch_2011, pyke_UnderstandingMovements_2015}. Reference \cite{sakiyama_LevylikeMovements_2016} shares results that seem to support this theory for Japanese carpenter ants (\textit{Camponotus japonicus}) in a featureless (circular) arena, though only when multiple ants are present, and only when considering the angular coordinate.

Other works have suggested that the turn angle distribution of a foraging or searching animal may be composed of a flat ``reorientation'' distribution, and a peaked (von Mises, wrapped Cauchy distribution, etc.) ``scanning'' distribution \cite{bartumeus_FractalReorientation_2008}. This decomposition may be seen as analogous to our decomposition of persistence into discrete tumbles and continuous rotational diffusion, as both approaches look to make a qualitative distinction between small and large angle changes. These are both useful ways to understand trajectory data, though one should be careful not to read too deeply into the behavioral implications of choosing one technique over the other. For example, while the applicability of the run-and-tumble model to the experimental data here allows for useful analysis of the trajectory statistics, we have no evidence that the ant's internal generation of this motion has any such discrete or continuous concepts. It is very possible that, to use the language previously introduced, the building blocks of the motion within an ant are vastly different from those presented here --- or potentially this generation mechanism can't even be said to be decomposable. Nonetheless, presenting different possible ways to generate these statistics gives us more ideas to eventually test against neural or other internally-recorded measurements that may be able to illuminate generative mechanisms.

An analytic expression for the persistence time near the boundaries is difficult to obtain for an arbitrary choice of boundary conditions. For simple, elastic reflecting boundary conditions (obtained by applying only half of the torque in Eq \ref{eq:torque_boundary}) we can expect that the persistence time will become limited by a term of the form $\sim L/v$, where $L$ is the characteristic size of the arena \cite{zamora_ExploringRunandtumble_2024}. In contrast, the aligning boundary conditions used here actually \textit{promote} persistence, instead of limiting it. It may be possible to derive an appropriate expression by considering a piecewise Fokker-Planck equation for a bounded domain \cite{fu_ConfinedRunandtumble_2024}, though that is beyond the scope of this work.

\section*{Conclusion}

We collected a large amount of data on the motion of \textit{A. gracilipes} in a finite arena without external stimuli, spanning nearly 100 individual specimens while controlling for various factors that might affect the ants' behavior. From these experiments, we extract the spatial position of the ants through time, allowing us to analyze their exploration behavior in this environment. We find that these trajectories show the characteristics of run-and-tumble-like motion, both qualitatively and quantitatively, by examining the distributions of run times, turn angles, and waiting times. We show that expressions for the velocity and persistence time derived under the assumptions of this model match closely with the statistics of the trajectories. Finally, we demonstrate that, among other advantages, the close agreement of the data with the proposed model allows for simulations of ant-like motion.

\section*{Supporting information}

\paragraph*{S1 Fig.}
\label{fig:S1}
\textbf{Results of control experiments on discretized statistics.} Discretized statistics for ant data split based on the three control groups described in the ``Ant tracking experiments'' section: whether the ants were isolated or kept together prior to tracking experiments, whether the arena was sanitized between tracking experiments, and whether the experiment was conducted in the morning or afternoon. All control/test pairs show essentially the same distributions and statistics as the pooled statistics (Fig 5, main text).

\paragraph*{S2 Fig.}
\label{fig:S2}
\textbf{Discretization benchmarking on simulated run-and-tumble data.} Estimated run-and-tumble parameters using the discretization scheme on simulated data compared to the prescribed parameter values using different discretization angles. Each dot represents a set of 80 simulated trajectories for 1200 seconds each, roughly the same quantity of data as in the experimental dataset. Black dashed line is $y=x$. Besides listed parameters, simulations used the following parameters: $dt = 1/60$s, $v = 25$mm/s, power law waiting time distribution with $\alpha = 1.75$. Top row shows results for an unbounded run-and-tumble, and the bottom row shows results for run-and-tumble simulated in the exact experimental geometry with all data within 10mm of the boundary removed. True persistence time is calculated using Eq 11 (main text). Error bars are standard deviations within each dataset for all values. 

\paragraph*{S3 Fig.}
\label{fig:S3}
\textbf{Turn angle, run time, and wait time distributions for various discretization parameter choices.} Statistical distributions for data shown in Fig 5 (main text) for different choices of angle and velocity parameters, $\theta_c$ and $v_c$. Top row distributions are color-coded based on the fit value from the lower plot. All statistics shown the same qualitative behavior, with only minor quantitative differences.

\paragraph*{S1 Appendix.}
\label{appendix:S1}
\textbf{Fokker-Planck formulation}

\paragraph*{S1 Table.}
\label{tab:S1}
\textbf{Parameters for SLEAP ant tracking models.}

\paragraph*{S2 Table.}
\label{tab:S2}
\textbf{Ant tracking trials that are excluded from analysis.}

\section*{Acknowledgments}

We are grateful for the equipment and support provided by the Engineering section of Core Facilities at OIST in creating the experimental arenas, as well as the computing resources provided by the Scientific Computing and Data Analysis section of Core Facilities at OIST.
We also thank the members of the Environmental Science and Informatics section at OIST, particularly Masako Ogasawara, Takumi Uchima, Toshihiro Kinjo, for their help identifying and collecting specimens.
Finally, we thank Samy Lakhal for particularly insightful discussions about analytics for the run-and-tumble model.


%
%

\bibliography{ant_tracking}

\includepdf[pages=-]{\supplementfilename}

\end{document}